\documentclass[aps,pre,superscriptaddress,twocolumn,showpacs]{revtex4-1}
\usepackage[utf8]{inputenc}
\usepackage{graphicx}
\usepackage{amsmath}
\usepackage[retainorgcmds]{IEEEtrantools}

\renewcommand{\d}{\text{d}}
\newcommand{\psib}{\overline{\psi}}
\newcommand{\avg}[1]{\left\langle#1\right\rangle}

\newlength{\figwidth}
\begin{document}

\title{Correlation functions of Ising spins on thin graphs} 

\author{Piotr Bialas}
\email{pbialas@th.if.uj.edu.pl}
\affiliation{Marian Smoluchowski Institute of Physics, Jagellonian University, Reymonta 4, 30--059 Krakow, Poland}
\affiliation{Mark Kac Complex Systems Research Centre, Faculty of Physics, Astronomy and Applied Computer Science,\\
  Jagellonian University, Reymonta 4, 30--059 Krakow, Poland}
\author{Andrzej K. Oleś}\email{oles@th.if.uj.edu.pl}
\affiliation{Marian Smoluchowski Institute of Physics, Jagellonian University, Reymonta 4, 30--059 Krakow, Poland}
\begin{abstract}
  We investigate analytically and numerically an Ising spin model with ferromagnetic coupling defined
  on random graphs corresponding to Feynman diagrams of a $\phi^q$ field theory, which exhibits a mean field phase transition. We explicitly calculate the correlation functions both in the symmetric
  and in the broken symmetry phase in the large volume limit.
They agree with the results for finite size systems obtained from
Monte Carlo simulations.
\end{abstract}

\maketitle

\section{Introduction}

The definition of a correlation function on random geometries is not
so obvious as in the usual fixed geometry case. The problem is that we
cannot just take two fixed points at a given distance because the
distance between them is changing. One possible solution is to
sum over all pairs of points at a given distance
\begin{equation}\label{def:GAB}
G^{AB}(r) \equiv \frac{1}{n}\avg{\sum_{ij}A_i B_j \delta_{d(i,j),r}}.
\end{equation}
Here $d(i,j)$ is the geodesic distance, {\em i.e.}, the shortest path
between points $i$ and $j$. $A$ and $B$ are some quantities defined in
each vertex whose correlation we want to measure. The average is taken
over all instances of the geometry (configurations) and $n$ is the size
of the system. Please note that
the distance $d(i,j)$ is a very nonlocal quantity: in principle it does
depend on the whole configuration not only on the endpoints $i$ and
$j$. Therefore the correlation \eqref{def:GAB} is not a two-point function,
which can lead to some interesting and non-intuitive behavior. 

Due to this non-locality the calculation of function
\eqref{def:GAB} is in general very difficult. As in other fields of statistical
physics an insight can be gained by examining simple solvable models,
the best known being probably the Ising model. In this paper we
study the Ising model on a random geometry ensemble made of all
Feynman diagrams of a $\phi^q$ theory.

The paper is organized as follows: 
in Sec.~\ref{sec:themodel} we present and analytically solve the model. 
We proceed on in Sec.~\ref{sec:correlations} calculating the spin-spin correlation functions using the similarity to Cayley trees in the infinite volume limit.
Section~\ref{influence} addresses the influence of spins on geometry.
Final discussion and summary are given in Sec.~\ref{summary}.

\section{The model}
\label{sec:themodel}

We consider an Ising ferromagnet spin model on $q$-regular random
graphs corresponding to $\phi^q$ Feynman diagrams of a
zero-dimensional field theory \cite{des,bachas}. The partition
function of this model is defined as a sum over all $\phi^q$ Feynman
diagrams $\mathcal{G}$ and all the values that the spins  $s_1, \ldots, s_n$ on a graph $G$ can take
\begin{equation}
Z=\sum_{G\in\mathcal{G}} \; \sum_{s_1, \dots, s_n} \! e^{-\beta H(G; s_1, \ldots, s_n )}.
\end{equation}
If no external
magnetic field is present the energy of the system on a single
 diagram $G$ reads
\begin{equation}\label{eq:H}
H(G; s_1, \ldots, s_n) = - \!\! \sum_{\langle i,j \rangle \in G} s_i s_j,
\end{equation}
where the sum is over all nearest neighbor pairs.  Two vertices are
considered as nearest neighbors if they are connected by a link. This
includes loops, in which case a vertex is its own
neighbor, and multiple links when two vertices are
counted multiple times in the sum \eqref{eq:H}.

Associating with the ``up'' and ``down'' spins the $\phi_{+}$ and
$\phi_{-}$ fields respectively, we can generate the requisite ensemble
from the Feynman diagram expansion of the partition function
\begin{equation}\label{eq:Z}
  Z = \int\! \d\phi_{+} \d\phi_{-} \text{exp} \left[
    -\frac{1}{2}\vec{\phi}^{\,T}\Delta^{\!-1}\vec{\phi}\; + \frac{1}{q!}
    \left( \phi_{+}^q + \phi_{-}^q \right) \right] ,
\end{equation}
where $\vec{\phi}^{\,T} \equiv (\phi_+,\phi_-)$  and $\Delta$ is the transfer
matrix
\begin{equation}\label{eq:delta}
 \Delta \equiv \begin{pmatrix}
 \ e^{\beta}  & \ e^{-\beta} \\
 \ e^{-\beta} & \ e^{\beta}
\end{pmatrix}.
\end{equation}
Following \cite{des} we define the coupling constant $g \equiv
e^{2\beta}$.

We will use the saddle-point approximation method to calculate the
partition function $Z$ in the large $n$ limit. We start by performing
binomial expansion of the $\phi^q$ terms in Eq.~\eqref{eq:Z}

\begin{IEEEeqnarray}{rCl}\label{eq:binomial}
  \exp \left[ \frac{1}{q!} \left( \phi_{+}^q + \phi_{-}^q
    \right) \right] & = &
  \sum_{n=0}^{\infty}\frac{1}{n!}\left[ \frac{1}{q!} \left(\phi_{+}^q + \phi_{-}^q \right) \right]^{n} \nonumber\\
  & = &\sum_{n=0}^{\infty} \frac{1}{(q!)^n} \sum_{k=0}^{n}
  \frac{\phi_{+}^{q k} \; \phi_{-}^{q
    (n-k)}}{k!  (n-k)!}  .
\end{IEEEeqnarray}
Substituting this back into Eq.~\eqref{eq:Z} and taking the term for a
specific $n$ only we get the partition function of an ensemble of
graphs with exactly $n$ vertices
\begin{equation}\label{eq:Zn}
Z_n =\frac{1}{(q!)^n}\sum_{k=0}^{n}C(k,n) \! \int \! d\phi_{+} d\phi_{-} \: e^{-\frac{1}{2}\vec{\phi}^{\,T}\Delta^{\!-1}\vec{\phi}} \: \phi_{+}^{q k} \: \phi_{-}^{q (n-k)}
\end{equation}
where
\begin{equation}\label{eq:C}
C(k,n) \equiv \frac{1}{k! (n-k)!}.
\end{equation}

At this point it is convenient to introduce the rescaled fields $\psi_\pm$ and
the variable $z$
\begin{displaymath}
\psi_{\pm} \equiv \frac{1}{\sqrt{n}}\;\phi_{\pm}, \quad z \equiv \frac{k}{n},
\end{displaymath}
and rewrite Eq.~\eqref{eq:Zn} as
\begin{equation}\label{eq:Zn2} 
 Z_n = \mathcal{C}(n) \int_{0}^{1} \!\! dz \, C(z,n) \int \! d\psi_{+} d\psi_{-} \: e^{n f(\psi_{+}, \psi_{-}) },
\end{equation}
where $\mathcal{C}(n)$ is a constant depending on $n$ only and 
\begin{equation}
f(\psi_{+}, \psi_{-}) = -\frac{1}{2}\vec{\psi}^{\,T}\Delta^{\!-1}\vec{\psi} + q \left[ z \ln{\psi_{+}} + (1\!-\!z) \ln{\psi_{-}} \right].
\end{equation}

The field integral in Eq.~\eqref{eq:Zn2} can be asymptotically
approximated using the saddle-point approximation
\begin{equation}
\int \! d\psi_{+} d\psi_{-} \; e^{n f(\psi_{+}, \psi_{-}) } \approx e^{n f(\psib_{+}, \psib_{-})} \; \frac{2\pi}{n\sqrt{\det \mathbf J}},
\end{equation}
where $\mathbf J$ is the Jacobian matrix $J_{xy} = f_{\psib_x\psib_y}$ . We use the shorthand notation
\begin{equation}
f_{\psib_x\psib_y} \equiv \left.\frac{\partial^2 f(\psi_+,\psi_-)}{\partial\psi_x\partial\psi_y}
\right|_{\psi_x=\psib_x,\psi_y=\psib_y}.
\end{equation} 
$\psib_\pm$ are given by the saddle-point equations
\begin{equation}
\frac{\partial f(\psib_+,\psib_-)}{\partial \psib_\pm}=0,
\end{equation}
which can be written in matrix form
\begin{equation}\label{eq:maximumcond}
\Delta^{\!-1} \binom{\psib_{+}}{\psib_{-}} = 
\binom{q \, z / \psib_{+}}{q \, (1-z) / \psib_{-}}.
\end{equation} 
The above system of two quadratic equations has in general four
solutions, from which only the one positive in both $\psi_{+}$ and
$\psi_{-}$ has physical meaning
\begin{IEEEeqnarray}{rCl}
\psib_{+}(z) & = & g^{-\frac{3}{4}}  \sqrt{\frac{q}{2}}\, \cdot \nonumber \\*[-.5em]
 && \cdot \left[ 1+ 2z \left(g^2-1\right) +\sqrt{1+4 \left(g^2-1\right) (1-z) z} \right]^{\frac{1}{2}} , \nonumber \\*[.5em]
\psib_{-}(z) & = & \psib_{+}(1-z) .
\end{IEEEeqnarray}
Because of Eq.~\eqref{eq:maximumcond} $f(\bar{\psi}_{+}, \bar{\psi}_{-})$
simplifies to
\begin{equation}\label{eq:fbar}
f(\bar{\psi}_{+}, \bar{\psi}_{-}) = q \left[ z \ln{\bar{\psi}_{+}} + (1-z) \bar{\psi}_{-} - \frac{1}{2} \right].
\end{equation}
The logarithm of the $C(z,n)$ term 
can be approximated using the Stirling formula
\begin{IEEEeqnarray}{rCl}\label{eq:logc}
\ln{C(z,n)}  & \approx &  -n \left[ z\ln{z} + (1-z)\ln{(1-z)}  \right] \nonumber \\
 && - \frac{1}{2} \ln{z(1-z)}  + \ln{(2\pi n)} + n\ln{n} - n. \nonumber \\
\end{IEEEeqnarray}
So finally 
\begin{equation}\label{eq:Zn3}
Z_n \approx \mathcal{C}(n) \int_0^1 \!\!\d{z}\, e^{n F(z)}
\end{equation}
and combining Eqs.~\eqref{eq:Zn2}, \eqref{eq:fbar}, and \eqref{eq:logc} we
obtain $F(z)$ to the leading order in $n$
\begin{IEEEeqnarray}{rCl}\label{eq:Fleading}
F(z) & \approx & - z\ln{z} - (1-z)\ln{(1-z)} \nonumber \\
 && +\: q \left[ z\ln{\bar{\psi}_{+}} + (1-z)\ln{\bar{\psi}_{-}} \right].
\end{IEEEeqnarray}
The next non-leading term is given in Appendix \ref{app:A}.

\begin{figure}
\begin{center}  
\includegraphics[width=\figwidth]{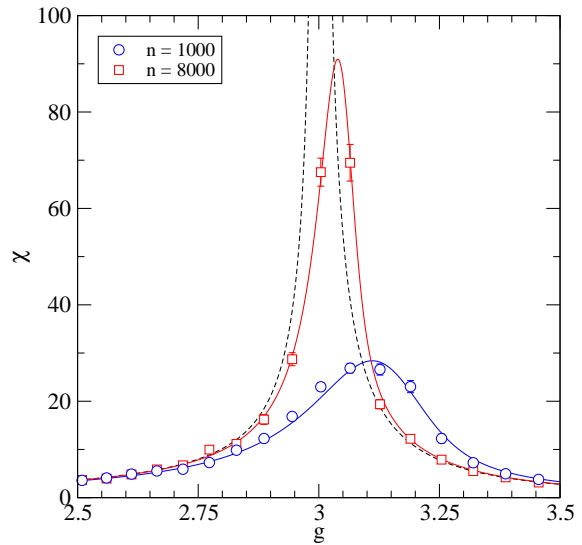}
\end{center}
\caption{\label{fig:M2}
$\tilde\chi$ as a function of the
  coupling $g$ for $q\!=\!3$. Circles (blue) render MC data for graphs of size $n\!=\!1000$ and squares (red) for those ones of size $n\!=\!8000$ vertices. The error bars show statistical uncertainty which increases near the transition. The solid lines plot the corresponding results of numerical integration whereas the saddle-point approximation is given by the dashed line.}
\end{figure}

Function $F(z)$ is symmetric around $z=1/2$ and in the disordered
phase it has a maximum at this point. Exactly at the transition
$F''(1/2)=0$, from which we can calculate the critical coupling
\begin{equation}\label{eq:gc}
g_c=\frac{q}{q-2}.
\end{equation}
For $g>g_c$ the $F(z)$ function has a minimum at $z=1/2$ and two maxima at $z
= 1/2(1 \pm m)$. In saddle-point approximation $m$ is the average
magnetization
\begin{equation}
\avg{M} \equiv 2 k -n = n(2 z-1) =n m .
\end{equation}  
Finding the zeros of the first derivative of $F(z)$ we get for $q=3$
\begin{equation}\label{eq:m3}
m=\frac{g}{g-2}\sqrt{\frac{g-3}{g+1}}
\end{equation}
and 
\begin{equation}\label{eq:m4}
m=\frac{g}{g^2-2}\sqrt{g^2-4}
\end{equation} in the $q=4$ case.
The susceptibility $\chi=\frac{1}{n}(\avg{M^2}-\avg{M}^2)$ can be
calculated in saddle-point approximation from the integral
\begin{align}\label{eq:iM2}
\avg{M^2}&=
n^2\frac{\int_0^1 \d{z}(2z-1)^2  \, e^{n F(z)}}
{\int_0^1 d{z}\, e^{n F(z)}},
\end{align}
leading to
\begin{equation}
  \chi=-\frac{4}{ F''\left(\frac{1}{2}(1+m)\right)} .
\end{equation}
In the symmetric phase where $m=0$ we obtain
\begin{equation}\label{eq:chi-form}
\chi=\left( 1-\frac{q}{2} \; \frac{g-1}{g} \right)^{\!\!-1} .
\end{equation}
In the broken symmetry phase we get for $q=3$
\begin{equation} 
\chi=\frac{4g}{(g-3)(g-2)^2(g+1)} .
\end{equation}
Actually, for comparison with Monte Carlo (MC) simulations we will use
\begin{equation} 
\tilde\chi=\frac{1}{n}(\avg{M^2}-\avg{|M|}^2)
\end{equation}
instead of $\chi$ because
average magnetization $\avg{M}$ is not well defined in numerical simulations. On a
finite lattice it is in principle zero for all values of
$\beta$. However, in the broken symmetry phase its measured value will
in general depend on the algorithm used and the duration of the
simulation. The average absolute value of magnetization is given
by the integral
\begin{align}\label{eq:iabsM}
\avg{|M|}=
n\frac{\int_0^1 \d{z}|(2z-1)|  \, e^{n F(z)}}
{\int_0^1 d{z}\, e^{n F(z)}}
\end{align}
and in the saddle-point approximation in the symmetric phase  it is equal to
\begin{equation}
\frac{1}{n}\avg{|M|}^2=\frac{2}{\pi}\chi .
\end{equation}
In the broken symmetry phase $\avg{|M|}=n\,m$. We plot the resulting
expression for $\tilde\chi$ (dashed line) together with data obtained from MC simulations in Fig.~\ref{fig:M2}.

Instead of doing saddle-point approximation of the integrals
\eqref{eq:iM2} and \eqref{eq:iabsM} we can integrate them numerically.
The results are plotted in Fig.~\ref{fig:M2} with solid lines. As one can see the agreement is 
very good already for small systems even if we keep the
leading order terms only given by Eq.~\eqref{eq:Fleading}. Including higher order
terms does not improve the result significantly.

\section{Correlations}
\label{sec:correlations}

In this section we will calculate the connected spin-spin correlation function
\begin{equation}\label{def:Gss-con}
G^{ss}_{c}(r)\equiv\frac{1}{n}\avg{\sum_{i,j}(s_i-m)(s_j-m)\delta_{d(i,j),r}}.
\end{equation}
This is not the only way to define the connected correlation
function on random geometry \cite{js}. We will use this one because it
has the usual property
\begin{equation}\label{eq:disp}
\sum_r G^{ss}_c(r)=\chi.
\end{equation}
Expanding the expression under the average we can express \eqref{def:Gss-con} as
\begin{IEEEeqnarray}{rCl}
  G^{ss}_{c}(r) & = & \frac{1}{n}\avg{\sum_{i,j}s_i s_j\delta_{d(i,j),r}} - 2m\frac{1}{n}\avg{\sum_{i,j}s_i\delta_{d(i,j),r}} \nonumber \\
&& +\: m^2\frac{1}{n}\avg{\sum_{i,j}\delta_{d(i,j),r}} \nonumber \\*[.5em]
& \equiv & G^{ss}(r)-2 m \, G^{s1}(r)+m^2 G^{11}(r).
\end{IEEEeqnarray}

To calculate the correlation functions we will use the fact that in the
leading order of $n$ the solution of the model coincides with the Bethe solution of
the Ising model on Cayley trees with coordination number equal to $q$
\cite{des}. Contrary to Cayley trees the Ising model on Feynman
diagrams studied here exhibits a genuine phase transition.

The $G^{11}(r)$ correlation function is a volume-factor: it is the
average number of vertices at the distance $r$ from a given vertex. 
On a Cayley tree with fixed geometry it is easy to calculate
\begin{equation}\label{eq:g11inf}
G^{11}(r)=q\,(q-1)^{r-1}.
\end{equation}
\begin{figure}
\begin{center}
\includegraphics[width=\figwidth]{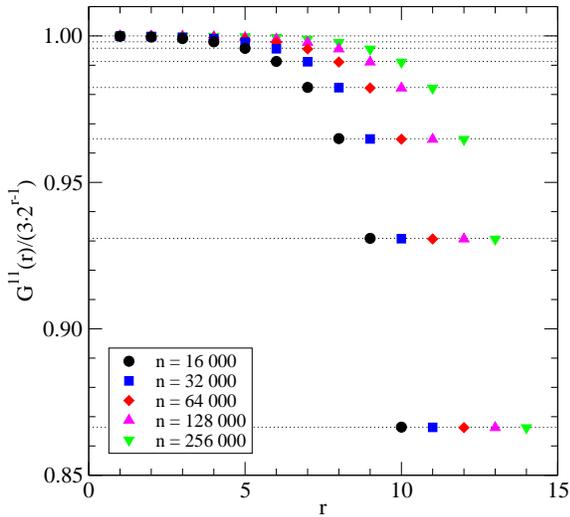}
\end{center}
\caption{\label{fig:g11r}
Ratio of the correlation function $G^{11}(r)$ obtained from MC simulations of ensembles of  various sizes with $q\!=\!3$ at $\beta\!=\!\beta_c\!=\!\frac{1}{2}\ln3$  to its infinite volume limit $q\,(q-1)^{r-1}$.}
\end{figure}
In Fig.~\ref{fig:g11r} we plot the ratio of the measured correlation function 
to the infinite volume limit \eqref{eq:g11inf}. Please note the scaling relation
evident from the plot
\begin{equation}
2G^{11}(r;n)=G^{11}(r+1;2n).
\end{equation}
This can be also written as
\begin{equation}
G^{11}(r)=n\,\mathcal{F}\left(\frac{2^r}{n}\right)=n\,\widetilde{\mathcal{F}}(r\ln2-\ln n).
\end{equation}
We have checked this relation and found that it is very well fulfilled already for graphs as small as 1000 vertices.

\begin{figure}
\begin{center} 
\includegraphics[width=7cm]{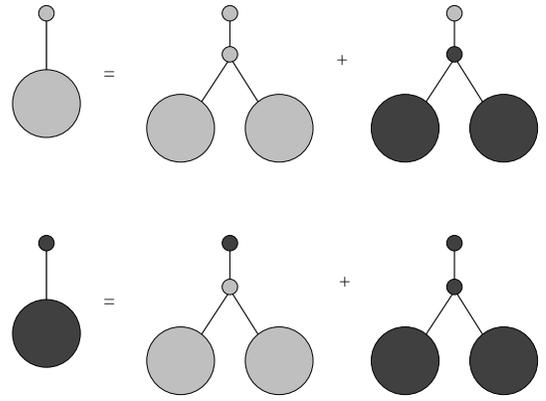}
\end{center}
\caption{\label{fig:saddle}Graphical representation of the saddle-point equations given by Eqs.~\eqref{eq:bochas} for $q\!=\!3$. The bright bubbles correspond to $\Phi_+$ and the dark ones to $\Phi_-$. Smaller bright and dark points stand for vertices carrying the ``up'' and ``down'' spins respectively. }
\end{figure}

The Ising model on Cayley trees is generated by the equations
\begin{equation}\label{eq:bochas}
\begin{cases}
\Phi_+&=\dfrac{1}{(q-1)!}\left(\sqrt{g}\,\Phi_+^{q-1}+\dfrac{1}{\sqrt{g}}\,\Phi_-^{q-1}\right) \\*[1em]
\Phi_-&=\dfrac{1}{(q-1)!}\left(\dfrac{1}{\sqrt{g}}\,\Phi_+^{q-1}+\sqrt{g}\,\Phi_-^{q-1}\right)
\end{cases}
\end{equation}
whose graphical interpretation is shown in Fig.~\ref{fig:saddle}.
\begin{figure}[t]
\includegraphics[width=7cm]{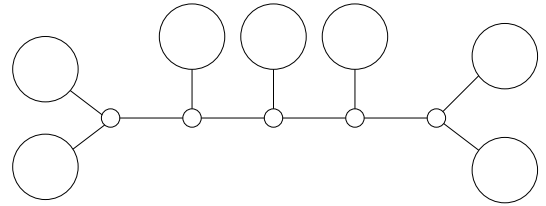}
\caption{\label{fig:correlation}Graphical representation of the
 correlation function $Z(r)_{xy}$ given by Eq.~\eqref{eq:zcor} for $q\!=\!3$. 
Bubbles depict $\Phi_\pm$ and points stand for individual vertices connected 
by links associated with $\Delta$. Only the first and the last vertex have
specific spins and each of them contributes by a $\frac{1}{2}\Phi_\pm^{2}$ factor.
 The spin of the $r\!-\!1$ vertices along the path connecting them
 can be either ``up'' or ``down'' which is covered by the matrix $\mathbf M$. }
\end{figure}
These equations are analogous to the ones obtained for Ising model on
branched polymers \cite{Ambjorn}. The difference is that here the
probability of ending the branch is zero. This formally means that we
have only infinite trees. Hence there is no chemical potential
associated with every vertex.

The partition function can be then obtained as
\begin{equation}\label{eq:Zbochas}
Z=\frac{1}{q!}\left(\Phi_+^q+\Phi_-^q\right).
\end{equation}
The system \eqref{eq:bochas} will have in general many  pairs of solutions.
However, for $g\le g_c$ the system is in symmetric phase and $\Phi_+=\Phi_-$. 
The one resulting equation  can be easily solved yielding
\begin{equation}\label{eq:phipm}
\Phi_\pm^{q-2}=\frac{\sqrt{g}}{1+g}\, (q-1)!=\frac{1}{2\cosh\beta}(q-1)! .
\end{equation}
When $g>g_c$  the dominant solution of Eqs.~\eqref{eq:bochas} will have $\Phi_+\neq\Phi_-$
giving a non-zero magnetization \cite{bachas}
\begin{equation}\label{eq:m-bochas}
m=\frac{\Phi_+^q-\Phi_-^q}{\Phi_+^q+\Phi_-^q}.
\end{equation}
Finding this solution requires solving the system \eqref{eq:bochas}. 
Although this cannot be done analytically for general $q$ for $q=3$ and $q=4$ we get in the broken phase
\begin{equation}\label{eq:q34}
\Phi_{\pm}=
\begin{cases}
\dfrac{\sqrt{g}}{g-1} \left(1 \pm \sqrt{\dfrac{g-3}{g+1}} \right) & \text{for $q=3$},\\*[1em]
\left[ \dfrac{3\sqrt{g} \left(g\pm\sqrt{g^2-4}\right)}{g^2-1} \right]^{\frac{1}{2}} & \text{for $q=4$}.
\end{cases}
\end{equation}
Substituting this into Eq.~\eqref{eq:m-bochas} we obtain the magnetization
which is identical to the expressions \eqref{eq:m3} and \eqref{eq:m4} calculated using the
previous method.

We will derive the correlation
functions by the method described in \cite{Bialas:2000xsa}. The
correlation function can be represented graphically as in Fig.~\ref{fig:correlation}.
To this picture corresponds the expression
\begin{equation}\label{eq:zcor}
Z(r)_{xy}=\frac{1}{\left[(q-1)!\right]^2}\Phi_x^{q-1} 
\left(\overbrace{\Delta \mathbf M\cdots\Delta \mathbf M}^{r-1} \Delta\right)_{\!\!\! xy}
 \!\! \Phi_y^{q-1}, 
\end{equation}
where 
\begin{equation}
\mathbf M \equiv \frac{1}{(q-2)!}\begin{pmatrix}
\Phi_+^{q-2} & 0 \\
0 & \Phi_-^{q-2}
\end{pmatrix}.
\end{equation}
If we rewrite matrix $\mathbf M$ as
\begin{equation}
\mathbf M=\frac{(\Phi_+\Phi_-)^{\frac{q-2}{2}}}{(q-2)!}\begin{pmatrix}
\left(\frac{\Phi_+}{\Phi_-}\right)^{\frac{q-2}{2}}& 0 \\
0 & \left(\frac{\Phi_-}{\Phi_+}\right)^{\frac{q-2}{2}}
\end{pmatrix}
\end{equation}
we can see that the correlation functions are related to the
correlation functions of the Ising spin chain in the effective
magnetic field $(q-2)h_{eff}$ with \cite{morita}
\begin{equation}
h_{eff} \equiv \frac{1}{2}\ln\frac{\Phi_+}{\Phi_-} .
\end{equation}
In the symmetric case in particular $h_{eff} = 0$.

The required correlation functions can be expressed using $Z(r)_{xy}$ as
\begin{align}
G^{11}(r)&=\frac{1}{Z} \sum_{xy} Z(r)_{xy},\\
G^{1s}(r)&=\frac{1}{Z} \sum_{xy} y Z(r)_{xy},\\
G^{ss}(r)&=\frac{1}{Z} \sum_{xy} x  y Z(r)_{xy}.
\end{align}

After introducing matrices
\begin{equation}
\mathbf M^{\frac{1}{2}}=\frac{1}{\sqrt{(q-2)!}}\begin{pmatrix}
\Phi_+^{\frac{q-2}{2}} & 0 \\
0 & \Phi_-^{\frac{q-2}{2}}
\end{pmatrix}
\end{equation}
and
\begin{equation}\label{eq:tildeq}
\widetilde \mathbf Q \equiv \mathbf M^{\frac{1}{2}}\Delta \mathbf M^{\frac{1}{2}}
\end{equation}
$Z(r)_{xy}$ given by Eq.~\eqref{eq:zcor} can be rewritten in the form
\begin{IEEEeqnarray}{rCl}
Z(r)_{xy}& = & (q-2)!\;\frac{\Phi_x^{q-1} M^{-\frac{1}{2}}_{xx}}{(q-1)!} 
\left(
\mathbf M^{\frac{1}{2}}\Delta \mathbf M^{\frac{1}{2}}
\right)^{\!r}_{\! xy}
 \frac{M^{-\frac{1}{2}}_{yy}\Phi_y^{q-1}}{(q-1)!} \nonumber \\
& = & (q-2)!\; \frac{\Phi_x^{\frac{q}{2}}}{(q-1)!} 
\;{\widetilde Q}^r_{xy}\;
 \frac{\Phi_y^{\frac{q}{2}}}{(q-1)!} .
\end{IEEEeqnarray}
Assuming that the eigenvalues of $\widetilde \mathbf Q$ are
$\lambda_1$ and $\lambda_2$, and that the corresponding normalized eigenvectors are
$(a,b)$ and $(-b,a)$ respectively, we can calculate the matrix power by
diagonalizing $\widetilde \mathbf Q$. 

After some algebra we obtain:
\begin{IEEEeqnarray}{rCl}\label{eq:cor}
G^{11}(r) & = & \frac{1}{Z}\; \frac{
 \lambda_1^r(a \Phi_+^{\frac{q}{2}}+  b \Phi_-^{\frac{q}{2}} )^2
+\lambda_2^r(b \Phi_+^{\frac{q}{2}}-a \Phi_-^{\frac{q}{2}} )^2
}{(q-1)!(q-1)}, \IEEEeqnarraynumspace \\*[.5em]
G^{1s}(r) & = & \frac{1}{Z}\; \frac{
 \lambda_1^r(a^2 \Phi_+^q-  b^2 \Phi_-^{q} )
+\lambda_2^r(b^2 \Phi_+^{q}-a^2 \Phi_-^{q} )
}{(q-1)!(q-1)}, \IEEEeqnarraynumspace \\*[.5em]
G^{ss}(r) & = & \frac{1}{Z}\; \frac{
 \lambda_1^r(a \Phi_+^{\frac{q}{2}}  -  b \Phi_-^{\frac{q}{2}} )^2
+\lambda_2^r(b \Phi_+^{\frac{q}{2}}  +  a \Phi_-^{\frac{q}{2}} )^2
}{(q-1)!(q-1)}. \IEEEeqnarraynumspace
\end{IEEEeqnarray}

In Appendix~\ref{sec:eigen} we show that 
\begin{IEEEeqnarray}{rCl}
\lambda_1 & = & q-1 \; ,\\*[.5em]
\lambda_2 & = & (q-1)\left[\frac{\sqrt{g}}{(q-1)!}(\Phi_+^{q-2}+\Phi_-^{q-2})-1\right] \nonumber \\*[.3em]
& = & (q-1)\left[\frac{2\sqrt{g}\,\Phi^{q-2}}{(q-1)!}\cosh\left( (q-2)h_{eff}\right) -1\right] \nonumber \\*[.3em]
& \equiv & (q-1) \; \widetilde\lambda_2 \; ,
\end{IEEEeqnarray}
where 
\begin{equation}
\Phi \equiv \sqrt{\Phi_+\,\Phi_-} .
\end{equation}
The corresponding eigenvector is equal to
\begin{equation}
(a,b)
=
\frac{(\Phi_+^{\frac{q}{2}} , \Phi_-^{\frac{q}{2}})}{\sqrt{\Phi_+^q+\Phi_-^q}}.
\end{equation}
Substituting this into Eq.~\eqref{eq:cor} we get 
\begin{equation}
G^{11}(r)=\frac{1}{Z}
\frac{ (q-1)^{r-1} }{(q-1)!}\;(\Phi_+^q+    \Phi_-^q ).
\end{equation}
Using the definition \eqref{eq:Zbochas} of $Z$ we finally obtain
the result \eqref{eq:g11inf} which is a check of the consistency of
the method used.

For spin-spin correlation functions we obtain
\begin{IEEEeqnarray}{rCl}\label{eq:s-s}
G^{1s}(r) & = & \frac{q!\,(q-1)^r}{(q-1)!\,(q-1)}\, \frac{\Phi_+^{q}-\Phi_-^{q}}{\Phi_+^{q}+\Phi_-^{q}} = m \; G^{11}(r)\,,\IEEEeqnarraynumspace \\*[.5em]
G^{ss}(r) & = & q!\,\frac{(q-1)^r(\Phi_+^q  - \Phi_-^q )^2 +4\Phi_+^q \Phi_-^q \lambda_2^r} {(q-1)!\,(q-1)(\Phi_+^q+    \Phi_-^q )^2} \nonumber \\*[.5em]
 & = & G^{11}(r) \left[ \frac{4\Phi_+^q \Phi_-^q}{(\Phi_+^q+    \Phi_-^q )^2} \widetilde\lambda_2^r\ + m^2 \right] \nonumber \\*[.5em]
 & = & G^{11}(r) \left( \frac{\widetilde\lambda_2^r}{\cosh\left(q\, h_{eff}\right)} + m^2 \right).
\end{IEEEeqnarray}
In the symmetric case when $m=0$ and $\Phi_+=\Phi_-$ are given by Eq.~\eqref{eq:phipm} we readily get
\begin{equation}\label{eq:gss31}
g^{ss}(r)\equiv\frac{G^{ss}(r)}{G^{11}(r)}=\left(\frac{g-1}{g+1}\right)^{\!r}=\tanh^r\beta.
\end{equation}
This is as predicted the result obtained for correlation of Ising spins on the
chain \cite{morita}.  It is easy to check that the
relation \eqref{eq:disp} is satisfied  by the above function.

\begin{figure}[!t]
\begin{center}
\includegraphics[width=\figwidth]{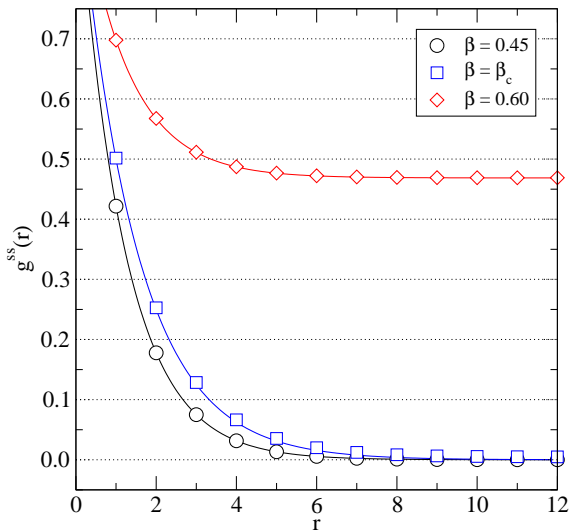}
\end{center}
\caption{\label{fig:corr}
Correlation function $g^{ss}(r)$ for $q\!=\!3$.  Symbols render MC data for graphs of size $n\!=\!256000$ in the
  symmetric phase for $\beta\!=\!0.45$ (circles), at the transition $\beta\!=\!\beta_c\!=\!\frac{1}{2}\ln3$ (squares), and in the broken phase
  for $\beta\!=\!0.60$ (diamonds). The solid lines plot the analytical predictions given by Eqs.~\eqref{eq:gss31} and \eqref{eq:gss32}. }
\end{figure}

In the broken phase inserting Eqs.~\eqref{eq:q34} into Eq.~\eqref{eq:s-s} we obtain 
for $q=3$ 
\begin{equation} \label{eq:gss32}
g^{ss}(r)=
\frac{4}{(g-2)^2(g+1)}\frac{1}{(g-1)^r}+m^2
\end{equation}
and for $q=4$
\begin{equation}
g^{ss}(r)
=\frac{4}{(g^2-2)^2}\frac{1}{(g^2-1)^r}+m^2 .
\end{equation}
Although similar formulas can be derived for higher values of $q$ they are much more complicated. 

Figure~\ref{fig:corr} plots the correlation function $g^{ss}(r)$
for \mbox{$q=3$} and different values of $\beta$. As one can see the agreement with MC results is very good.
\begin{figure}
\begin{center}
\includegraphics[width=\figwidth]{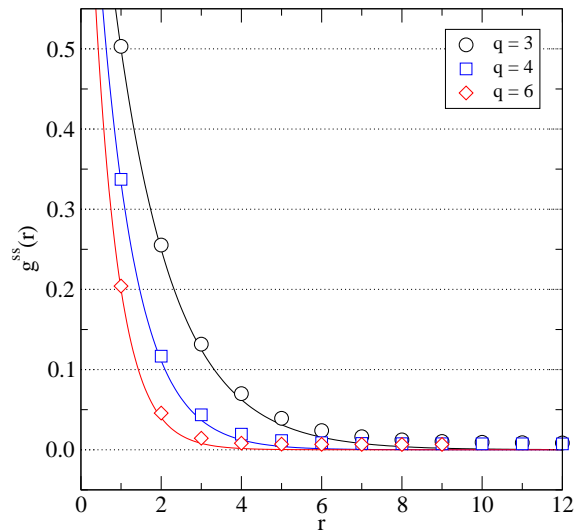}
\end{center}
\caption{\label{fig:fig6} 
Correlation function $g^{ss}(r)$ at the transition $\beta\!=\!\beta_c(q)$ for different values of $q$. Points represent MC results for graphs of size $n\!=\!64000$ while the solid lines plot the analytical predictions. Please note that for $q\!=\!6$ the maximum encountered distance (the diameter of the graph) is $r\!=\!9$.}
\end{figure}
In Fig.~\ref{fig:fig6} correlation functions for
different $q$'s at the respective transition points $\beta_c(q)=\frac{1}{2}\ln\frac{q}{q-2}$ are compared. Again, MC results match the asymptotic predictions. The small discrepancies
diminish with the increase of the graphs' size.

\section{Influence of spins on geometry}
\label{influence}

\begin{figure}
\begin{center}
\includegraphics[width=\figwidth]{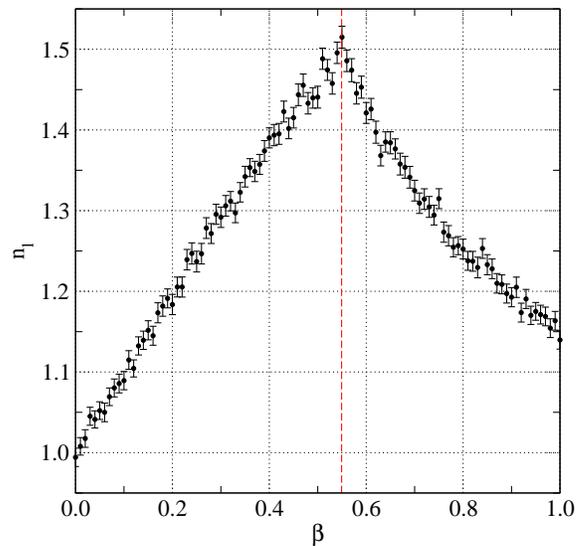}
\caption{\label{fig:loop}
Average number of loops $n_l$ in graphs of
  size $n\!=\!32000$ vertices and $q\!=\!3$. The black points represent MC data and the dashed red line marks the transition point $\beta\!=\!\beta_c$.}
\end{center}
\end{figure}
On a finite tree spins can be integrated out exactly and the resulting
factor does not depend on the shape of the tree. This means that the
spins do not have any influence on the geometry. The situation changes
when cycles are allowed, which is especially obvious for loops, {\em i.e.}, 
links attached at both ends to the same vertex. Without any
spins or $\beta=0$ one can show combinatorially that the
expected number of such loops on a graph equals $(q-1)/2$. For positive $\beta$,
however, we should observe an enhancement as each loop contributes
the $e^\beta$ factor to the partition function contrary to links joining
different vertices which can have different spins. For
$\beta\rightarrow\infty$ all the spins have the same sign and again
the geometry decouples. 

In Fig.~\ref{fig:loop} we plot the average number of loops $n_l$ as
a function of $\beta$. The results agree qualitatively with the above
scenario. Nevertheless, one should note that while looking pronounced this is
still only a $1/n$ effect. The number of loops is independent of $n$
and negligible in the large volume limit.

\section{Summary and discussion}
\label{summary}

We have analyzed in detail a simple model of spins on a random
geometry. In the large $n$ limit it is formally equivalent to the
Ising model on an infinite Cayley tree. However, it is well defined,
has a genuine phase transition, and can be easily simulated using
Monte Carlo methods.

We have derived expressions for correlation functions in both the 
symmetric and the broken phase with methods less formal then in
\cite{morita}. From these calculations we obtain a picture of the
transition. The correlation function $g^{ss}(r)$ does not exhibit any
critical behavior: the correlation length is finite for any finite
$\beta$. Nevertheless, the volume factor $G^{11}(r)$ grows exponentially
and this growth offsets the decay of the correlation function $g^{ss}(r)$.
As we increase $r$ by one, the influence of the spins at this distance
drops by $(g-1)/(g+1)$, however, the number of these spins increases by
$q-1$.  When
\begin{equation}
\frac{g+1}{g-1}=q-1
\end{equation}
all shells of spins contribute equally and we observe the phase
transition.  It is easy to check that this gives the expression
\eqref{eq:gc} for the critical value of $g$.

\acknowledgments{} We would like to thank Zdzisław Burda for many
helpful discussions. This research was supported in part by the PL-Grid Infrastructure. MC simulations were performed on the Shiva computing
cluster at the Faculty of Physics, Astronomy and Applied Computer
Science, Jagellonian University, and at the Academic Computer Centre CYFRONET AGH using the Zeus cluster.
\appendix

\section{Next to leading corrections to $F(z)$}

\label{app:A}
The Jacobian matrix $J_{xy} = f_{\psib_x\psib_y}$ is equal to
\begin{equation} 
\mathbf J = \Delta^{\!-1}-q\begin{pmatrix}
 z/\psib_+ & 0\\
0 &  (1-z)/\psib_-
\end{pmatrix},
\end{equation}
from which 
\begin{equation}
\det \mathbf J = \frac{g}{g^2-1} \left[1+q\,\sqrt{g}\left(\frac{z}{\psib_+^{\,2}}+ \frac{1\!-\!z}{\psib_-^{\,2}} \right) \right] + q^2\frac{z\,(1\!-\!z)}{\left(\psib_-\psib_+\right)^2}.
\end{equation}
and the approximation of the function $F(Z)$ given by Eq.~\eqref{eq:Fleading} takes the form
\begin{IEEEeqnarray}{rCl}
    F(z) & \approx & - \left[ z\ln{z} + (1-z)\ln{(1-z)} \right] \nonumber\\*[.2em]
    && +\: q \left[ z\ln{\bar{\psi}_{+}} + (1-z)\ln{\bar{\psi}_{-}} \right] \nonumber\\*[.1em]
    && -\: \tfrac{1}{2n}\left\{ \ln{(\det \mathbf J)}  +\ln{\left[ z\,(1-z) \right]} \right\}.
\end{IEEEeqnarray}

\section{Eigenvalues}
\label{sec:eigen}

Equations \eqref{eq:bochas} can be rewritten as
\begin{IEEEeqnarray}{rCl}
\begin{pmatrix}
\Phi_+ \\
\Phi_-
\end{pmatrix}& = & \frac{1}{(q-1)!}
\begin{pmatrix}
\sqrt{g}\Phi_+^{q-2} & \frac{1}{\sqrt{g}}\Phi_-^{q-2}\\
\frac{1}{\sqrt{g}}\Phi_+^{q-2} & \sqrt{g}\Phi_-^{q-2}
\end{pmatrix}
\begin{pmatrix}
\Phi_+ \\
\Phi_-
\end{pmatrix}\nonumber\\
& \equiv & \mathbf Q\cdot\begin{pmatrix}
\Phi_+ \\
\Phi_-
\end{pmatrix}.
\end{IEEEeqnarray}
The above equation means that matrix $\mathbf Q$ has an eigenvalue equal
to one and that the corresponding eigenvector is proportional to
$(\Phi_+,\Phi_-)$.

Matrix $\widetilde \mathbf Q$ defined in Eq.~\eqref{eq:tildeq} can be expressed as
\begin{equation}
\widetilde Q_{xy}=(q-1) \: Q_{xy}\left(\frac{\Phi_x}{\Phi_y}\right)^{\!\! \frac{q-2}{2}}.
\end{equation}
One can check that this implies that $\widetilde \mathbf Q$ has an eigenvalue
$\lambda_1=q-1$ and the corresponding normalized eigenvector
\begin{equation} 
(a,b)=\frac{(\Phi_+^{\frac{q}{2}},\Phi_-^{\frac{q}{2}})}{\sqrt{\Phi_+^{q}+\Phi_-^{q}}} .
\end{equation}
The second eigenvector is perpendicular to this one. Multiplying it by
$\widetilde \mathbf Q$ we obtain the second eigenvalue.

\end{document}